\documentclass[nopreprintline,3p,times]{elsarticle}

\usepackage{graphicx}
\usepackage{amssymb}
\usepackage{bbm}
\RequirePackage{amssymb,amsmath,bm}
\usepackage{subfigure}
\usepackage{lineno}

\begin{document}

\begin{frontmatter}


\title{Optimization of the Area Under the ROC Curve using Neural Network Supervectors for Text-Dependent Speaker Verification}


\author{Victoria Mingote, Antonio Miguel, Alfonso Ortega, Eduardo Lleida}

\address{
  ViVoLab, Arag\'{o}n Institute for Engineering Research (I3A), University of Zaragoza, Spain
  }
\address{\{vmingote,amiguel,ortega,lleida\}@unizar.es}

\begin{abstract}
This paper explores two techniques to improve the performance of text-dependent speaker verification systems based on deep neural networks.
Firstly, we propose a general alignment mechanism to keep the temporal structure of each phrase and obtain a supervector with the speaker and phrase information, since both are relevant for a text-dependent verification.
As we show, it is possible to use different alignment techniques to replace the global average pooling providing significant gains in performance.
Moreover, we also present a novel back-end approach to train a neural network for detection tasks by optimizing the Area Under the Curve (AUC) as an alternative to the usual triplet loss function, so the system is end-to-end, with a cost function close to our desired measure of performance.
As we can see in the experimental section, this approach improves the system performance, since our triplet neural network based on an approximation of the AUC (\emph{aAUC}) learns how to discriminate between pairs of examples from the same identity and pairs of different identities.
The different alignment techniques to produce supervectors in addition to the new back-end approach were tested on the RSR2015-Part I database for text-dependent speaker verification, providing competitive results compared to similar size networks using the global average pooling to extract supervectors and using a simple back-end or triplet loss training.
\end{abstract}


\end{frontmatter}


\section{Introduction}
\label{S:1}

The performance in many speaker verification tasks has improved thanks to deep learning advances in signal representations  \cite{Taigman2014DeepFace:Verification,snyder2018x} or optimization metrics \cite{Hoffer2015DeepNetwork,schroff2015facenet,snyder2016deep} that have been adapted from the state-of-the-art deep learning face verification systems.
In this paper, we propose alternatives to the two following aspects.
First, the generation of signal representations or embeddings which preserve the subject identity and uttered phrase.
Second, a metric for training the system which, as we will show, is more appropriate for a detection task.

In many recent verification systems, deep neural networks (DNNs) are trained for multi-class classification.
A common approach is to apply a global average reduction mechanism \cite{Taigman2014DeepFace:Verification,snyder2018x,Malykh2017OnTask}, which produces a vector representing the whole utterance which is called embedding.
A simple back-end as a similarity metric can be applied directly for the verification process \cite{nguyen2010cosine} or more sophisticated methods like the one presented in \cite{li2018angular}. 
However, this approach does not work efficiently in text-dependent tasks since the uttered phrase is a relevant piece of information to correctly determine the identities and the system has to detect a match in the speaker and the phrase to be correct \cite{Malykh2017OnTask,Liu2015DeepVerification}.
In our previous work \cite{Mingote2018}, we have noted that part of the imprecisions may be derived from the use of the average as a representation of the utterance, and how this problem can be solved by adding a new internal layer into the deep neural network architecture which uses an alignment method to encode the temporal structure of the phrase in a supervector.
In this paper, we propose a generalization for the use of different alignment mechanisms that can be employed in combination with the deep neural network to generate a differentiable supervector with good performances as will be shown in the experiments. 

The second proposal of the paper is an alternative to the triplet loss optimization, initially proposed in face recognition \cite{Hoffer2015DeepNetwork,schroff2015facenet}.
The goal of the learning process using triplet loss is to maximize the similarity metric or minimize the distance between two examples that belong to the same identity while minimizing the similarity or maximizing the separation margin with the third example from a different identity.  
We propose the use of a novel method to combine the triplet philosophy with a  loss function which approximates the Area Under the Receiver Operating Characteristic (ROC) Curve \cite{garcia2012optimization,toh2008maximizing,herschtal2004optimising,soriano2015fusion} in a differentiable way.
The Area Under the Curve (AUC) measures the area between the ROC and the axes, and the AUC is also a performance measure independent of the operating point.
Therefore, by maximizing its value we can improve the overall performance, as we will show in the paper. 

In the context of text-independent speaker verification tasks, the combination of i-vector extraction and Probabilistic Linear Discriminant Analysis (PLDA) \cite{kenny2008study,dehak2011front} is still dominant. 
%
%
Nonetheless, recently, several components have progressively been replaced by DNNs.
%
Examples of this are the use of DNN bottleneck features instead of spectral parametrization or combined with it \cite{lozano2016analysis}, the use of phonetic posteriors obtained by DNN acoustic models for alignment instead of Gaussian Mixture Model (GMM) in i-vector extractors \cite{lei2014novel}, or replacing PLDA by a DNN \cite{ghahabi2014deep}.
More ambitious proposals have been proposed to train DNNs for speaker classification with a large number of speakers as classes, and then extract embeddings of an intermediate layer by reduction mechanisms which are also called x-vectors \cite{snyder2018x,bhattacharya2017deep,snyder2017deep}.
%

The application of the previous text-independent techniques for text-dependent speaker verification tasks has produced mixed results. 
Traditional techniques with specific modifications have achieved good performance for these tasks such as i-vector+PLDA \cite{zeinali2017hmm}, DNN bottleneck as features for i-vector extractors \cite{zeinali2016deep}, posterior probabilities for i-vector extractors \cite{zeinali2016deep,dey2016deep} or using the speaker and channel latent factor mechanism \cite{Miguel2014FactorRecognition} with autoencoder DNNs \cite{Miguel2017TiedRecognition}.
On the other hand, when the task involves a large amount of data and only one phrase, good results have been provided by speaker embeddings obtained directly from a DNN \cite{Heigold2016End-to-endVerification}.
While, in tasks with more than one phrase and smaller databases, the lack of data may lead to problems with the use of deep architectures due to overfitting.
For this reason, these techniques have been shown ineffective \cite{Malykh2017OnTask,Liu2015DeepVerification}.

In our previous work \cite{Mingote2018}, we explored another reason for the lack of effectiveness in these tasks.
The order of the phonetic information of the uttered phrase is relevant for the identification.
Most of the approaches to obtain speaker embeddings from an utterance reduce temporal information, so this kind of systems only maintain the information of who is speaking and they may not capture the phonetic information for the final identification process.
For that reason, we integrated a new layer in the DNN architecture as an alignment mechanism based on Hidden Markov Model (HMM) \cite{rabiner1989tutorial}.
However, this alignment technique has some limitations, so in this work, we explore other alternative technique to make the alignment process.
%

%

In summary, in this paper, we propose a new optimization procedure that combines the triplet loss with an approximation of the AUC (\emph{aAUC}) as metric to maximize the inter-speaker similarity and minimize the intra-speaker similarity simultaneously. 
Furthermore, we remark the relevance of keeping speaker and phrase information to achieve a correct verification process, and propose a generalization of the differentiable alignment mechanism using a new alignment technique based on GMM posteriors that creates the supervector, providing better performance on the RSR2015 dataset compared to previous similar approaches \cite{Malykh2017OnTask,Liu2015DeepVerification}.

This paper is organized as follows. In Section 2 we describe our system and the different alignment strategies developed. Section 3 presents the triplet network method based on a novel triplet loss function. Section 4 presents the experimental setup. In Section 5 we explain the results achieved. Conclusions are presented in Section 6.

\section{Deep Neural Network based on Alignment}
To encode the phrase and speaker information from the audio file into a representation vector, we have developed a differentiable alignment mechanism for neural networks.
%
This representation has a common mechanism with the supervector in speaker verification.
It has the advantage of being discriminative against differences in the phonetic information, which is needed in our task, and, at the same time, it is robust to other sources of variability like the speed or the way the utterance is pronounced, even by the same person.

%
%
%
%

\subsection{Architecture}
Deep speaker verification systems are usually composed by three main parts: Front-End (FE), Pooling (Pool), and Back-End (BE).
We refer as Front-end to the first part of the systems that obtain the acoustic features from the input signal, and usually apply a Convolutional Neural Network (CNN) \cite{Malykh2017OnTask} or a Time Delay Neural Network (TDNN) \cite{novoselov2018deep} to the acoustic features to increase the temporal context of the representation.
The variable time length signals that are processed by the Front End are converted into fixed length vectors in a Pooling part of the system. 
This process usually consists of an average reduction of the temporal dimension of the input signal, but we have recently proposed a mapping between the input signal and the state components of an alignment \cite{Mingote2018}. 
The output of the Pooling part of the system will be an embedding or supervector.
These embeddings with fixed dimension are processed by the
Back-end to train a classification of identities, therefore, enforcing speaker separability in the embedding space or to provide directly the verification scores as we propose later in the paper.

\begin{figure}[th]
    \centering
	\subfigure[Architecture type A]{
 	\includegraphics[width=0.235\linewidth]{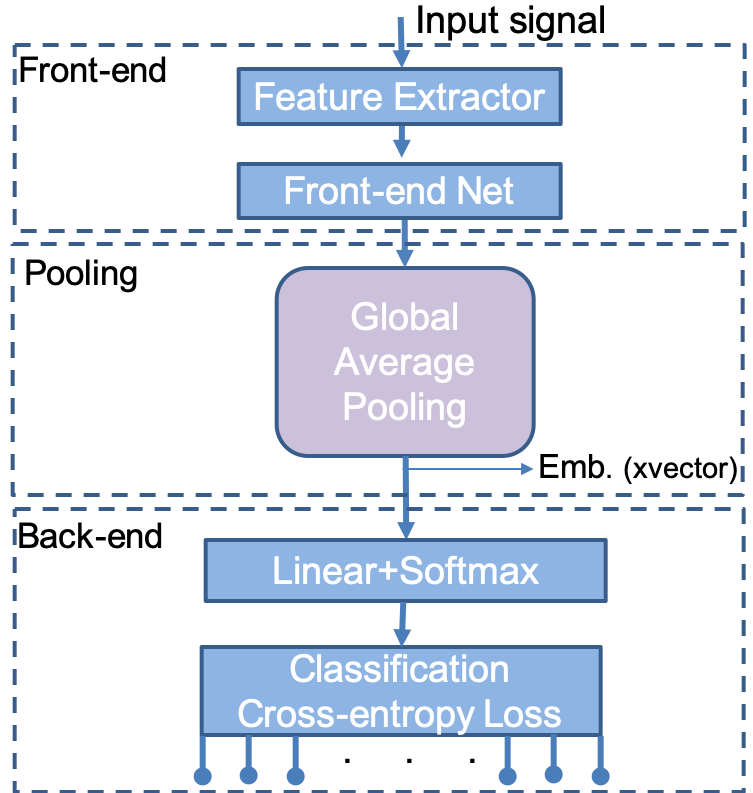}
	\label{fig1:a}}
	\centering
	\subfigure[Architecture type B]{
 	\includegraphics[width=0.235\linewidth]{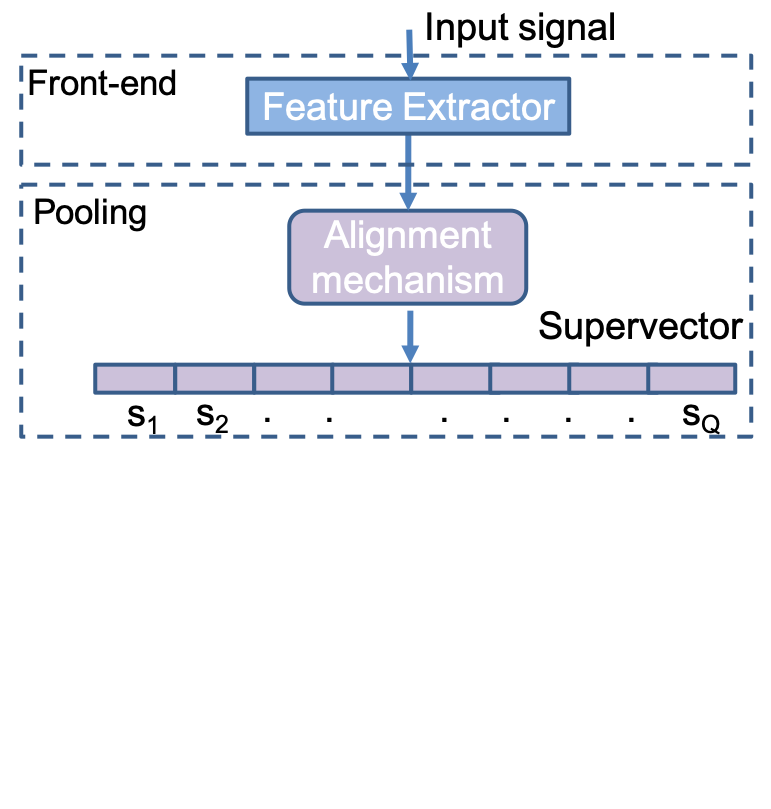}
	\label{fig1:b}}
    \centering
	\subfigure[Architecture type C]{
 	\includegraphics[width=0.235\linewidth]{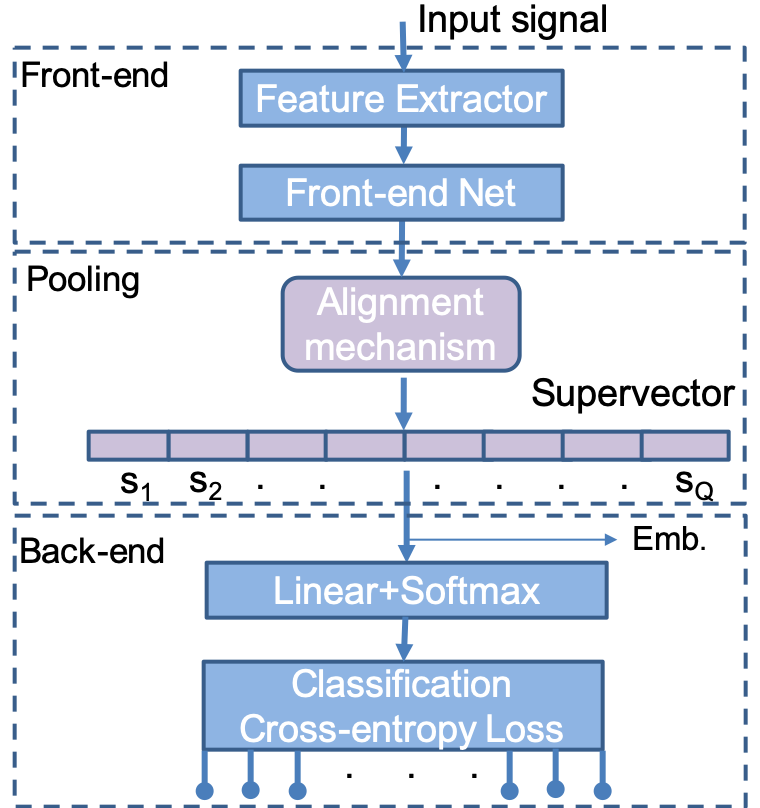}
	\label{fig1:c}}
    \centering
	\subfigure[Architecture type D]{
	\includegraphics[width=0.235\linewidth]{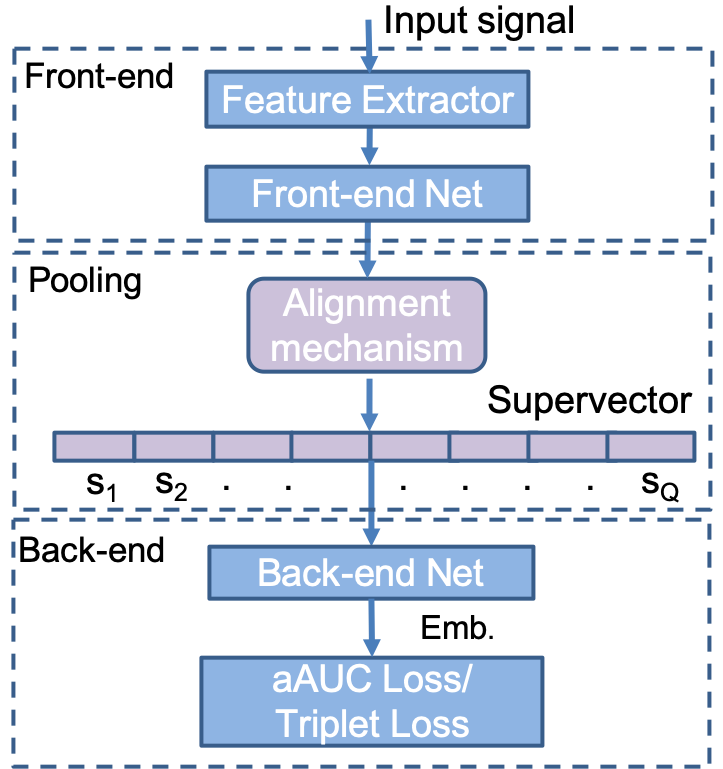}
    \label{fig1:d}}
    \caption{The architectures used to develop this work, \ref{fig1:a} the architecture type A is trained for multiclass classification using a traditional global average pooling mechanism. In \ref{fig1:b} the architecture type B, the acoustic features are aligned directly to obtain the supervector. \ref{fig1:c} the architecture type C is trained for multiclass classification and used as pre-train for the other architecture. \ref{fig1:d} the architecture type D is trained to optimize the back-end net.} 
    \label{fig1}
\end{figure}

In Fig.~\ref{fig1}, we show the architectures employed to develop this work. 
These architectures are designed as follow:
\begin{itemize}
    \item \textit{Architecture A}. The architecture depicted in Fig.\ref{fig1:a} is the approach used in many recent verification systems, where a DNN is trained for multi-class classification, and a global average pooling is applied to extract the embedding representation.
    Once the system is trained, one embedding is extracted for each enroll and test file, and then a cosine similarity metric is applied over them to obtain the verification scores. 
    
    \item \textit{Architecture B}. As we showed \cite{Mingote2018}, the previous architecture with this kind of pooling is not the most suitable approach for text-dependent speaker verification tasks. 
    For this reason, we substitute the global average pooling by the differentiable alignment mechanism which allows us to keep the temporal structure of the utterance.
    In this work, the alignment techniques used are an HMM as well as a GMM combined with a Maximum a Posteriori (MAP) adaptation. 
    As the phrase transcription is known in text-dependent tasks, we could construct a specific left-to-right HMM model or a different GMM for each phrase of the data.
    Thus, we use the system in Fig.\ref{fig1:b} to check that these alignment mechanisms work better than the global average polling.
    %
    This architecture consists of applying the alignment mechanism as a layer directly on the acoustic features, so we obtain the traditional supervector.
    
    \item \textit{Architecture C}. In view of the good behaviour of the alignment layer and the possibility to propagate gradients, we can use this layer as a link between front-end and back-end parts which allows us to train the whole system to optimize any cost function we decide.
    Thus, we create the architecture depicted in Fig.\ref{fig1:c} adding to the architecture type B a front-end network, and we train the architecture with a Cross-Entropy loss function.
    
    \item \textit{Architecture D}. In our previous work \cite{Mingote2018}, we employed only the architectures with a classification cross-entropy as cost function, but in this paper, as we describe in next section, we propose another architecture with a novel cost function in this context.
    This function is better integrated with the back-end and, therefore, close to the objective of the task.
    In the architecture type D which we show in Fig.\ref{fig1:d}, one embedding is obtained for each utterance, and then the back-end is applied to provide the verification scores directly with the metric which allows us to have an end-to-end system.
\end{itemize}

\subsection{CNN}
As front-end network, we have used a straightforward CNN architecture with only a few 1-dimension convolution (1D convolution) layers. 
These layers allow us to operate in the temporal dimension to add context information to the process, and at the same time, all the channels are combined at each layer.

To train efficiently this kind of layer, all the files should have the same length, and it possible to obtain this fixed length by applying an interpolation transformation or padding with zeros the input signals.
We have used an interpolation to have all of them with the same dimensions $(D^{0} \times T)$ where $D^{0}$ is the acoustic feature dimension, and $T$ is the temporal dimension.
Then, the input signal and its context, the previous and the subsequent frames indicated by the kernel size, are processed by the 1D convolution layer, where they are multiplied frame by frame with the corresponding weight matrix to produce a new vector sequence. 
The result of this operation is that each frame and its context is linearly combined to create the output signal.
During the training process with these convolution layers, the value for the temporal dimension ($T$) is kept until the pooling block, while the acoustic feature dimension changes when it passes through the network layers in function of the channel combination of each layer, $D=(D^{(0)},D^{(1)},...,D^{(L)})$ where $L$ is the total number of front-end layers. 

%

\subsection{HMM Alignment Mechanism}
Firstly, in our previous work \cite{Mingote2018}, we developed our experiments only with a phrase HMM-based alignment mechanism.
Using this approach, the knowledge of the phonetic information is not needed in the training process, so we can train easily an independent HMM model for each different phrase. 
Furthermore, this kind of alignment mechanism has a left-to-right architecture which employs the Viterbi algorithm to provide a decoded sequence in which all the HMM states have correspondence to at least one frame of the utterance, so no state is empty.

The process followed to add this frame-to-state alignment to our system is detailed below.
We obtained a sequence of decoded states \textbf{$\overline{q}$}=$(q_{1},...,q_{t},...,q_{T})$ where \textit{$q_{t}$} indicates the decoded state at time t with $\textit{$q_{t}$}\in\{1, ..., Q\}$ from the HMM models and \textbf{$Q$} is the number of states. 
After that, these decoded sequence vectors are converted into a matrix with ones at each state according to the frames that belong to this state and zeros in the rest of states, so we have the alignment matrix $\textit{A}\in\mathbb{R}^{T \times Q}$ with its components $\textit{$a_{tq_{t}}$=1}$ and $\textit{$\sum_{q}a_{tq}$=1}$ which means that only one state is active at the same time.

Once this process is over, the alignment matrix is provided externally to the network for each utterance providing the supervector as output of a matrix product.
This makes easier to differentiate and enables to backpropagate gradients to train neural network as usual.
The matrix multiplication assigns the sum of the corresponding frames to each state vector, resulting in a supervector if we consider them concatenated.
This process can be expressed as a function of the input, \textbf{$x^{L}_{dt}$}, to this layer $L$ with dimensions $(D^{(L)} \times T)$ and matrix of alignment of each utterance \textbf{$A$} with dimensions $(T \times Q)$:
\begin{equation}
x^{(L+1)}_{dq}=\frac{\sum_{t}x^{(L)}_{dt}\cdot a_{tq}}{\sum_{t}a_{tq}},
\end{equation}
where $x^{(L+1)}_{dq}$ is the supervector of the layer $(L+1)$ with dimensions $(D^{(L)} \times Q)$, where there are \textit{Q} state vectors of dimension $D^{(L)}$ and we normalize with the number of times state \textit{q} is active.

\subsection{GMM with MAP Alignment Mechanism}
Now, we propose the use of a GMM to generalize the previous method.
This method provides more flexibility since a single frame might not correspond to only one component in the mixture, they can be distributed, and there might be empty components for the whole sequence.

The philosophy of this frame-to-components alignment process is similar to the one described in the previous section, but in this case, we obtain the GMM alignment from the posterior probability of the hidden variables.
Matrix \textbf{$A$} with dimensions \textit{(T $\times$ C)} where \textit{C} are the components of the GMM is built by assigning the posterior probabilities of Gaussian component $c$ at time $t$ to the elements of the matrix $a_{tc} = \gamma_t(c) = P(Z=c | \mathbf{x}^{(L)}_{dt})$.
Often times most of the posteriors are close to zero.
To prevent the loss of performance due to the sparness of matrix \textbf{$A$} we add a MAP adaptation \cite{reynolds1995robust,reynolds2000speaker}. 
To define how this layer operates, we employ the following expression:
\begin{equation}
x^{(L+1)}_{dc}=\frac{\sum_{t} x^{(L)}_{dt} \cdot \gamma_t(c) +{\tau\cdot\mu_{dc}^{(b)}}}{\sum_{t}\gamma_t(c)+\tau} ,
\end{equation}
where $x^{(L+1)}_{dc}$ is the supervector of the layer $(L+1)$, $\tau>0$ is the relevance factor, $\gamma_t(c)$ is the posterior probability of Gaussian component $c$, $x^{(L)}_{dt}$ is the input to this layer $L$, and $\mu_{dc}^{(b)}$ is the running mean per component of the mixture which will be updated each batch $b$ in a similar manner to a Batch Normalization layer \cite{Ioffe2015BatchShift}: 
\begin{equation}
\label{eqn:mu}
\mu_{dc}^{(b)}=(1-\beta)\cdot\mu_{dc}^{(b-1)} + \beta\cdot{f},
\end{equation}
where $\beta$ is the adaptation coefficient and f is the mean estimation using the batch data samples.

The MAP adaptation ensures that the components with a low count of activations in a sequence will be assigned the mean value of the corresponding supervector section, $\mu_{dc}^{(b)}$, making the system more robust.  

\section{Triplet Neural Network}
The triplet neural network, referred in Fig.\ref{fig1} as back-end, defines a cost function to evaluate the embeddings provided by three instances of the same neural network with shared parameters.
As input of this network three examples are used, an example from a specific identity \textbf{$e$} (an anchor), another example from the same identity \textbf{$e^+$} (a positive example), and an example from another identity \textbf{$e^-$} (a negative example).
In most of the existing systems using this approach \cite{snyder2016deep,zhang2017end,dey2018end,novoselov2018triplet}, the network architecture has been trained with the triplet loss function \cite{snyder2016deep} which maximizes the distance between the anchor and the negative example at the same time that the distance between the anchor and the positive example is minimized if it is greater than a margin.
%
Unlike the previous systems, we decided to use a loss function which is more intuitive for the detection task, since this function allows us to optimize directly the AUC metric that measures the performance of our whole system.

Furthermore, another important point to train correctly this kind of systems is the triplet data selection applied to choose which are the examples that compose the triplets. 
We decided to use a similar approach to the triplet sampling strategy proposed in \cite{schroff2015facenet} which is usually called Hard Negative Mining instead of a random selection.
This technique consists in selecting the anchor-negative pairs with the maximum similarity value (hard negative) for which the system triggers a false alarm, and the anchor-positive pairs with the minimum similarity value (hard positive) which the system can not detect and produces a miss.

\begin{figure}[th]
  \centering
  \includegraphics[width=0.75\linewidth]{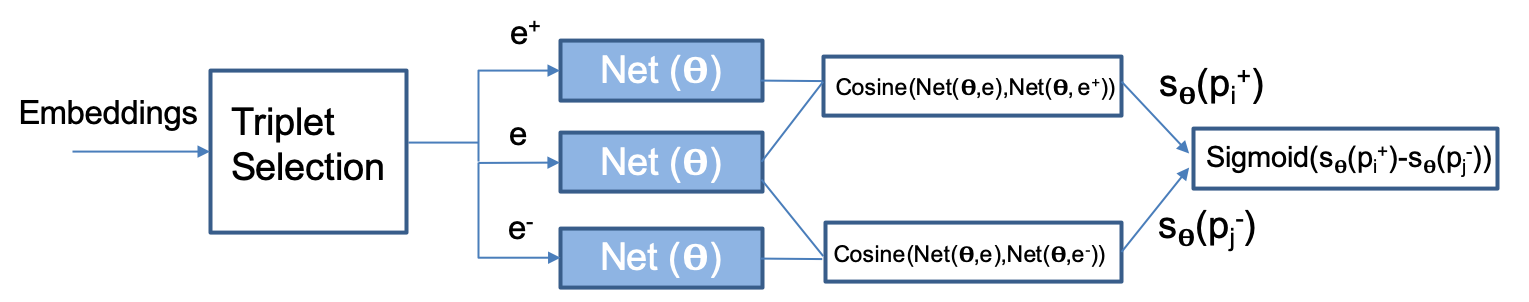}
  \caption{Triplet neural network, the embeddings are grouped in triplets by the triplet selection to train the network and evaluated the two pairs of embeddings to optimize the objective function.} 
  \label{fig4}
\end{figure}

The pipeline of the proposed scheme for training our triplet neural network back-end is depicted in Fig.\ref{fig4}.

\subsection{Optimization of the Area Under the ROC Curve}
Verification systems are generally trained with discriminative paradigms to optimize the classification performance.
However, in that way their training process does not consider the trade-off between false alarms and misses.
For that reason, we propose to optimize directly the AUC which is an operating point independent metric and measures the probability that pairs of examples are ranked correctly.
Since this metric is not differentiable, we propose an effective approximation as we show in the experiments.

For \textbf{$m$} training examples, the AUC is defined as the function that maximizes the average number of times the score for the anchor-positive pairs is greater than the score of the anchor-negative pairs.
The anchor-positive score is given by the cosine similarity value $s_{\Theta}(p^+_{i})$ where $p^+_{i}=(e,e^+)$ indicates each pair of anchor-positive embeddings with $\textit{$i$}\in\{1, ..., m^+\}$ and $m^{+}$ is the total number of positive examples, and the anchor-negative score is provided by the cosine similarity value $s_{\Theta}(p^-_{j})$ where $p^-_{j}=(e,e^-)$ represents the anchor-negative pairs with $\textit{$j$}\in\{1, ..., m^-\}$ and $m^{-}$ is the total number of negative examples.
Both values are also expressed as a function of the net learning parameters $\Theta$.
Therefore, given a set of network parameters $\Theta$, the AUC function can be written as  
%
\begin{equation}
AUC(\Theta)=\frac{1}{{m^+}{m^-}}{\sum_{i}^{m^+}\sum_{j}^{m^-}\mathbbm{1}(s_{\Theta}(p^+_{i})>s_{\Theta}(p^-_{j}))},
\end{equation}
where $\mathbbm{1}()$ has a value equal to '1' whenever $s_{\Theta}(p^+_{i})>s_{\Theta}(p^-_{j})$, and '0' otherwise. This function can be rewritten using the unit step function as, 
%
\begin{equation}
AUC(\Theta)=\frac{1}{{m^+}{m^-}}{\sum_{i}^{m^+}\sum_{j}^{m^-}u(s_{\Theta}(p^+_{i})-s_{\Theta}(p^-_{j}))}.
\label{eq:aucreal}
\end{equation}
%
%
%
To enable the backpropagation of the gradients, this expression must be approximated in order to be differentiable.
For that reason, we substitute the step function by a sigmoid function, so our approximation of the AUC (\emph{aAUC}) loss function can be formulated as,
%
\begin{equation}
aAUC(\Theta)=\frac{1}{{m^+}{m^-}}{\sum_{i}^{m^+}\sum_{j}^{m^-}{\sigma(\alpha\dot(s_{\Theta}(p^+_{i})-s_{\Theta}(p^-_{j})}))},
\label{eq:aucappr}
\end{equation}
%
where $\sigma()$ is the sigmoid function, and $\alpha$ is an adjustable parameter which modifies the slope of the sigmoid to make it more similar to the unit step \cite{soriano2015fusion}.
Thus, we seek the network parameters $\Theta^{*}$ which maximize this expression:
\begin{equation}
\Theta^{*}=\operatorname*{argmax}_\Theta\frac{1}{{m^+}{m^-}}{\sum_{i}^{m^+}\sum_{j}^{m^-}{\sigma(\alpha\dot(s_{\Theta}(p^+_{i})-s_{\Theta}(p^-_{j})}))}.
\end{equation}
%
%
\section{Experimental Setup}
\subsection{Data}
The experiments have been reported on the RSR2015 text-dependent speaker verification dataset \cite{Larcher2014Text-dependentRSR2015}.
This dataset comprises recordings from 157 male and 143 female.
For each speaker, there are 9 sessions with 30 different phrases.
This data is divided into three speaker subsets: background (bkg), development (dev) and evaluation (eval). 
We develop our experiments with Part I and we employ the bkg and dev data (194 speakers, 94 female/100 male) for training. The evaluation part is used for enrollment and trial evaluation. 

In this work, we have used this dataset for the experiments development since it is the only one composed by several phrases, and with train and test sets publicly available for text-dependent phrase based speaker verification.
Moreover, this dataset has three evaluation conditions, but in this work, we have only evaluated the most challenging condition which is the Impostor-Correct case where the non-target speakers pronounces the same phrase as the target speakers.

\subsection{System Configuration}
To develop our experiments, a set of features composed of 20 Mel-Frequency Cepstral Coefficients (MFCC) \cite{mermelstein1976distance,davis1980comparison} with their first and second derivates are employed as input to train the alignment mechanisms and also as input to the DNN.
The bkg partition has been employed to train two different alignment mechanisms, both were trained to obtain one model per phrase without the needed of knowing the phrase transcription.
On the one hand, HMM models have been trained using a left-to-right model of 40 states for each phrase. 
On the other hand, a 64 componenent GMM has been trained per phrase. 
With these models, we can extract the alignment information to use inside our DNN architecture. 
The lack of data may lead to DNN training problems, so we try to avoid a possible overfitting in our models with a data augmentation method called Random Erasing \cite{zhong2017random} which is applied on the input features.
Furthermore, in this work, we train the neural networks using an Adam optimizer \cite{kingma2014adam}. 
In all the experiments with the \emph{aAUC} function, the $\alpha$ parameter in the sigmoid function has a fixed value of 10 to have a shape close to the unit step.

\section{Results and Discussion}
In this work, two sets of experiments have been developed.
First, we focus the experiments on the different approaches for the front-end and the alignment mechanisms.
The best front-end neural network for each alignment mechanism is used as the reference in the second experiment set.
This second part of experiments compares the diverse back-end loss functions.

\subsection{Results of the Front-end Approaches}
As the first set of experiments, we compare architecture \emph{A} which makes use of a global average pooling (avg) \cite{Malykh2017OnTask,Liu2015DeepVerification}, architecture \emph{B} which applies the alignment to the feature input directly, and the proposed alignments combined with the front-end network using architecture \emph{C}. 
We employed two different neural network alternatives for the front-end network of architecture \emph{C}, both alternatives and the global average pooling approach are trained with a Cross-Entropy (CE) loss function and tested with a cosine similarity on the embeddings. 
The first one is a convolutional network with three layers (CNN) and a kernel of dimension 3 which gave us the best performance in \cite{Mingote2018}.
The second one replicates this basic network using a teacher and student version, following the Bayesian Dark Knowledge approach (bdk) \cite{balan2015bayesian}.
With this philosophy, a teacher network produces soft speaker identity labels which are used to train a student network.
In our case, Random Erasing data augmentation is applied to the input of the teacher network which adds variability to the labels prediction.
The student network is trained using these labels, and we are able to better capture the data variability introduced.
Thus, this architecture is used to provide robustness to the training of the neural network front-end.
%

In Table \ref{tab:table1}, we can see the Equal Error Rate (EER), the NIST 2010 minimum detection costs (\emph{DCF10} \footnote{https://www.nist.gov/sites/default/files/documents/itl/iad/mig/NIST$\_$SRE10$\_$evalplan-r6.pdf.}), and the AUC results for these experiments. 
As we showed, the embeddings from the global average reduction mechanism do not represent correctly the relevant information of the speaker and the phrase due to the importance of keeping the temporal structure of the phrases in the supervectors with the alignment layer within the DNN architecture.
Besides, applying the \emph{CNN} using the \emph{bdk} approach for the front-end network in architecture \emph{A}, the results are still worse. 
%
We can also observe that architecture \emph{C} which combines a front-end network with the alignment mechanisms improves the performance achieved applying only the alignment mechanism as in architecture \emph{B}. 
Furthermore, architecture \emph{C} using the \emph{CNN(bdk)} approach combined with the proposed frame-to-components alignment mechanism based on GMM with MAP provides an additional performance improvement.

\begin{table}[th]
  	\caption{Experimental results on RSR2015 part I \cite{Larcher2014Text-dependentRSR2015} eval set, showing AUC\%, EER\% and NIST 2010 min cost (\emph{DCF10}). These results were obtained to compare the global average pooling networks and the different neural networks with both alignment techniques.}
  	\vspace{0.2cm}
  	\label{tab:table1}
  	\centering
  	\resizebox{0.90\textwidth}{!} {
  	\begin{tabular}{c c c c c c c}
    \hline    
    \multicolumn{4}{c}{\textbf{Architecture}}&
    \multicolumn{3}{c}{\textbf{Results (EER\%/DCF10/AUC\%)}}\\
    \cline{1-4}
    \multicolumn{1}{c}{\textbf{Type}}&
    \multicolumn{1}{c}{\textbf{FE}}&
    \multicolumn{1}{c}{\textbf{Pool.}}&
    \multicolumn{1}{c}{\textbf{BE}}&
    \multicolumn{1}{c}{\textbf{Fem}}& 
    \multicolumn{1}{c}{\textbf{Male}}& 
    \multicolumn{1}{c}{\textbf{Fem+Male}}\\
    \hline
    $A$&$CNN$&$avg$&$CE$& $9.11/0.96/97.04$ & $8.66/0.96/97.21$& $8.88/0.96/97.13$\\
    $ $&$CNN(bdk)$&$ $&$ $& $30.48/1.0/76.81 $ & $31.48/1.0/75.45 $& $31.02/1.0/76.12$\\
    \cline{1-7}
    $B$&$-$&$HMM$&$-$&$1.34/0.21/99.80$&$1.57/0.16/99.77$&$2.03/0.29/99.71$\\
    \cline{2-7}
    $ $&$-$&$GMM$&$-$&$1.44/0.24/99.79$&$1.55/0.24/99.77$& $1.74/0.28/99.77$\\
    \cline{1-7}
    $C$&$CNN$&$HMM$&$CE$&\textbf{0.59}/\textbf{0.10}/$\textbf{99.95}$&$\textbf{0.71}/0.16/\textbf{99.96}$&$\textbf{0.73}/\textbf{0.14}/\textbf{99.95}$\\
    $ $&$CNN(bdk)$&$ $&$ $&$0.73/0.12/\textbf{99.95}$&$0.79/\textbf{0.14}/99.94$& $0.80/0.15/\textbf{99.95}$\\
    \cline{2-7}
    $ $&$CNN$&$GMM$&$CE$& $0.79/0.17/99.95$  & $0.99/0.23/99.94$& $0.92/0.20/99.94$\\
    $ $&$CNN(bdk)$&$ $&$ $& $\textbf{0.51}/\textbf{0.12}/\textbf{99.98}$  & $\textbf{0.78}/\textbf{0.15}/\textbf{99.96}$& $\textbf{0.66}/\textbf{0.13}/\textbf{99.97}$\\   
    \hline
	\end{tabular}}
\end{table}

\begin{figure}[th]
	\centering
 	\includegraphics[width=0.45\linewidth]{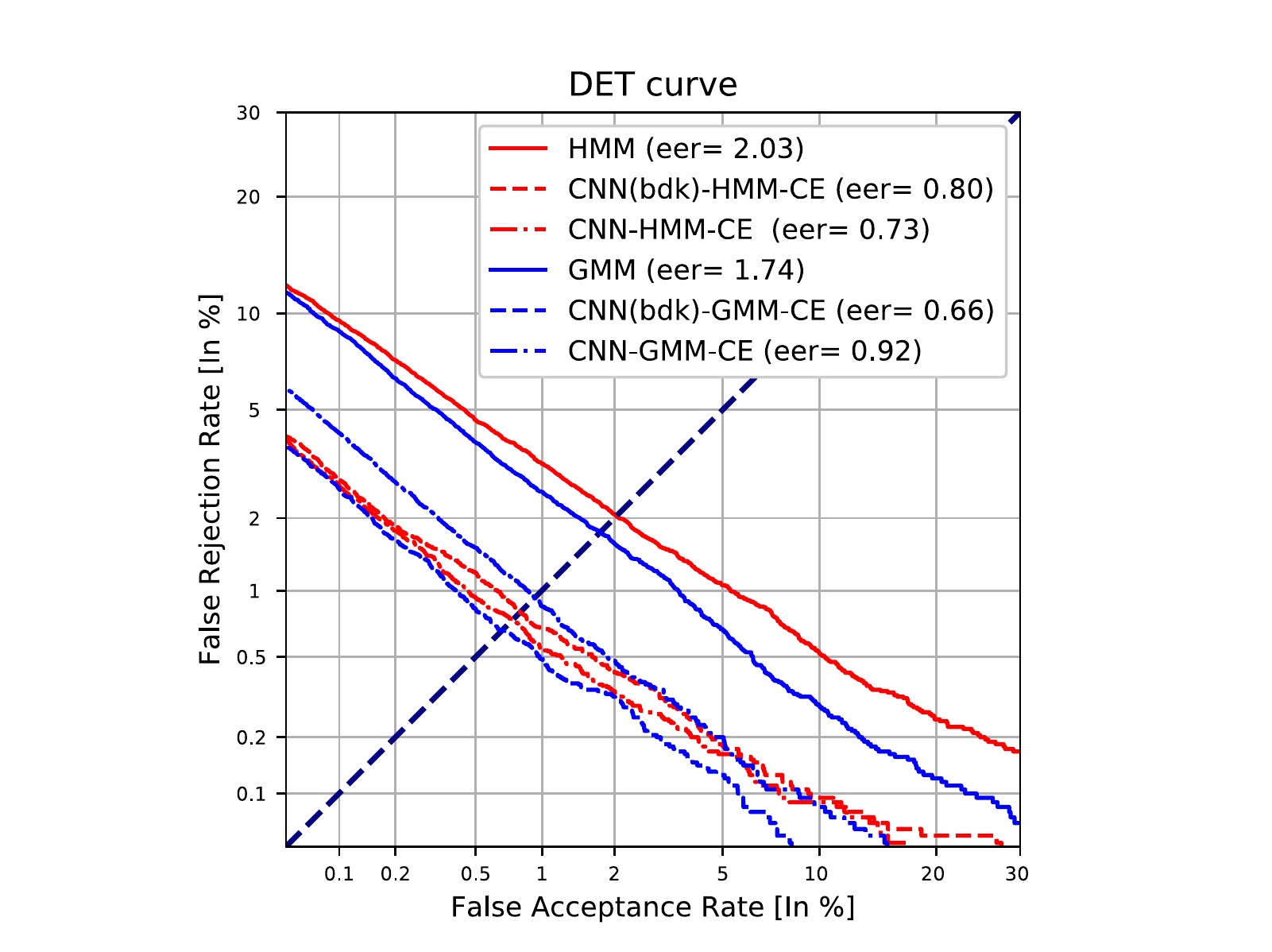}
    \caption{DET curves for female+male results of the different approaches employed for the front-end and alignment mechanism.} 
    \label{fig5}
\end{figure}

In addition to the previous table, we depict the Detection Error Trade-off (DET) curves in Fig.\ref{fig5}.
These curves show the results for female+male experiments. 
The global average pooling experiments are not represented since they do not provide relevant information.
These representations demonstrate that the approach with the best system performance is architecture \emph{C} with the front-end based on bdk and GMM with MAP as alignment mechanism.

\subsection{Results of the Back-end Approaches}
In the second experiment set, we contrast the best architecture \emph{C} for each alignment mechanism with architecture \emph{D} which is composed of two dense layers as back-end network and cosine similarity as metric, and we also compare the proposed \emph{aAUC} optimization to the triplet loss (TrLoss).
To initialize the second architecture, we employ a pre-trained model with architecture \emph{C}.

Table \ref{tab:table2} shows that if we apply the network back-end to the supervectors instead of using only the cosine similarity, we improve the ability to discriminate between different identities and phrases.
This is due to the joint estimation of the front-end and back-end network parameters to optimize the \emph{aAUC} objective.
Therefore, this is an end-to-end system since all the parameters are learned to jointly optimize the detection task.
The best system performance is obtained when we use the end-to-end \emph{aAUC} optimization strategy to combine the front-end and back-end networks with the alignment mechanism, which plays the important role of encoding the features into the supervector communicating both networks.
%
%
However, we achieve $11\%$ and $15\%$ relative improvement of EER$\%$ using HMM and GMM with MAP respectively, and the minDCF is also improved in both cases.  

\begin{table}[th]
  	\caption{Experimental results on RSR2015 part I \cite{Larcher2014Text-dependentRSR2015} eval set, showing AUC\%, EER\% and NIST 2010 min cost (\emph{DCF10}). These results were obtained to compare the best neural network front-end with both alignment techniques using the different back-ends.}
  	\vspace{0.4cm}
  	\label{tab:table2}
  	\centering
  	\resizebox{0.90\textwidth}{!} {
  	\begin{tabular}{c c c c c c c}
    \hline    
    \multicolumn{4}{c}{\textbf{Architecture}}&
    \multicolumn{3}{c}{\textbf{Results (EER\%/DCF10/AUC\%)}}\\
    \cline{1-4}
    \multicolumn{1}{c}{\textbf{Type}}&
    \multicolumn{1}{c}{\textbf{FE}}&
    \multicolumn{1}{c}{\textbf{Pool.}}&
    \multicolumn{1}{c}{\textbf{BE}}&
    \multicolumn{1}{c}{\textbf{Fem}}& 
    \multicolumn{1}{c}{\textbf{Male}}& 
    \multicolumn{1}{c}{\textbf{Fem+Male}}\\
    \hline
    \cline{1-7}
    $C$&$CNN$&$HMM$&$CE$&$0.59/0.10/99.95$&$0.71/0.16/99.96$& $0.73/0.14/99.95$\\
    $D$&$ $&$ $&$TrLoss$&$0.89/0.23/99.94$&$1.09/0.24/99.93$&$1.07/0.26/99.93$\\
    $ $&$ $&$ $&$aAUC$&$\textbf{0.52}/\textbf{0.10}/\textbf{99.96}$&$\textbf{0.67}/\textbf{0.14}/\textbf{99.97}$&$\textbf{0.65}/\textbf{0.13}/\textbf{99.96}$\\
    \cline{1-7}
    $C$&$CNN(bdk)$&$GMM$&$CE$& $0.51/0.12/\textbf{99.98}$  & $0.78/0.15/99.96$& $0.66/0.13/99.97$\\   
    $D$&$ $&$ $&$TrLoss$& $0.63/0.14/99.96$  & $0.70/0.16/\textbf{99.97}$& $0.72/0.17/99.96$\\ 
    $ $&$ $&$ $&$aAUC$&$\textbf{0.40}/\textbf{0.09}/\textbf{99.98}$& $\textbf{0.69}/\textbf{0.14}/\textbf{99.97}$& $\textbf{0.56}/\textbf{0.12}/\textbf{99.98}$\\
    \hline
	\end{tabular}}
\end{table}

\begin{figure}[th]
    \centering
	\includegraphics[width=0.45\linewidth]{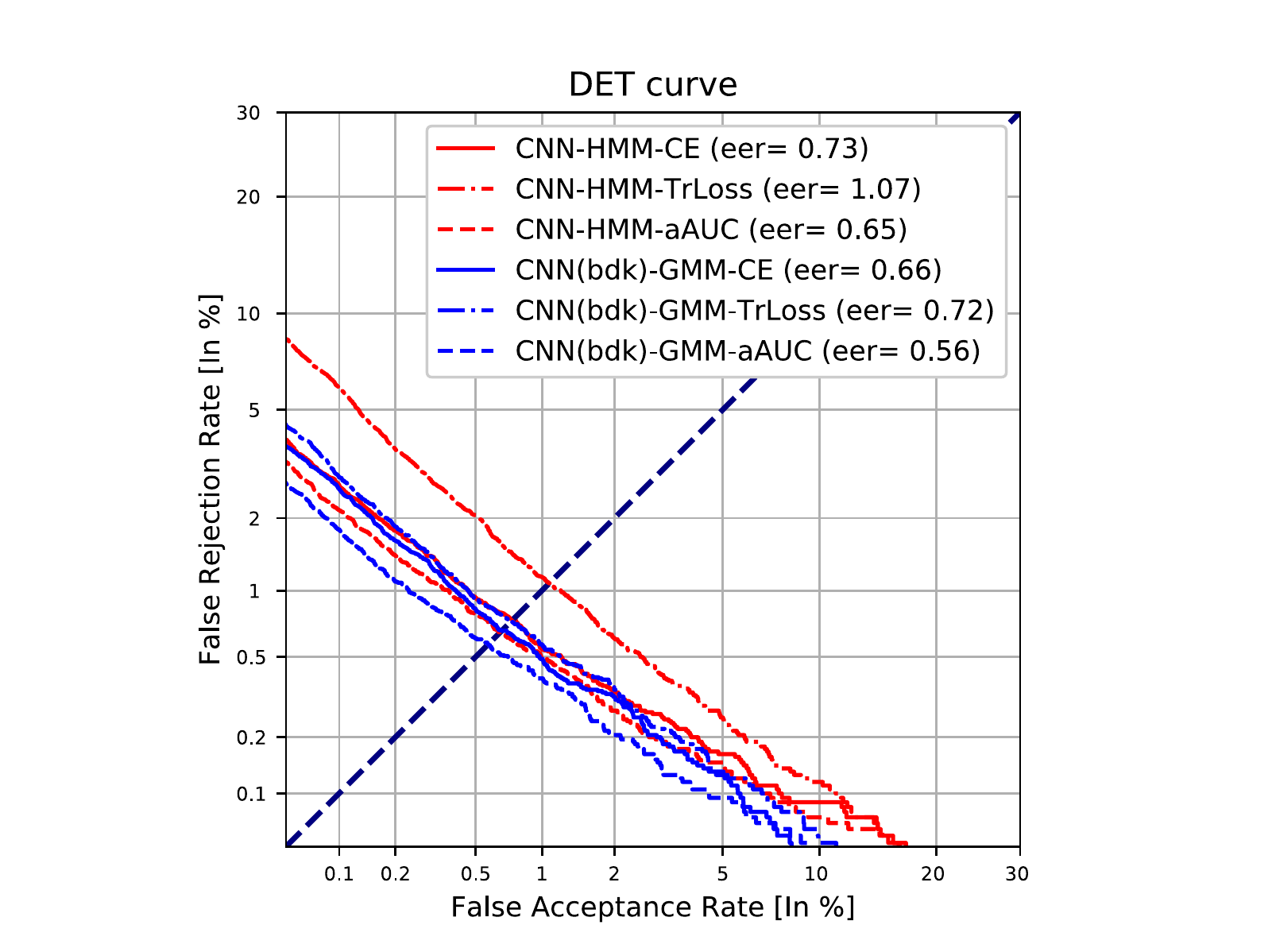}
    \caption{DET curves for female+male results of the best front-end networks combined with the different back-ends.} 
    \label{fig6}
\end{figure}

In Fig.\ref{fig6}, we represent the DET curves for these experiments.
These representations clearly demonstrate that the systems with GMM with MAP as alignment mechanism have a great system performance with all the approaches, and especially when we apply the proposed \emph{aAUC} function for the back-end, the best system performance is achieved.

\begin{figure}[th]
  \centering
  \includegraphics[width=0.45\linewidth]{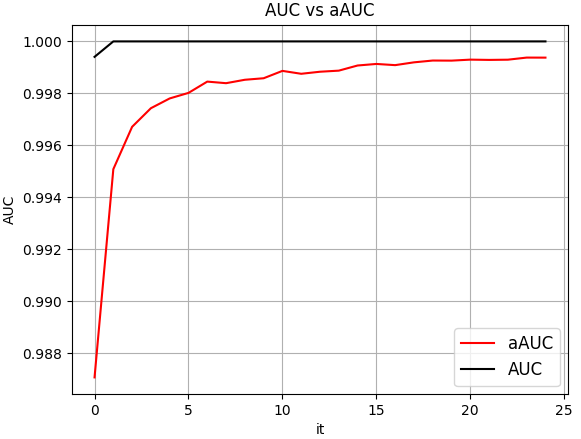}
  \caption{Training evolution of real AUC vs aAUC.} 
  \label{fig7}
\end{figure}

For illustrative purposes, Fig.\ref{fig7} represents the evolution of the real AUC function \eqref{eq:aucreal} against the \emph{aAUC} proposed \eqref{eq:aucappr} during the training process.
In this representation, we can see that the proposed differentiable estimation of the AUC function is getting close to the real function as the training progresses, which supports the assertion that the \emph{aAUC} is an effective approximation of the real AUC function.

\vspace{-0.2cm}
\section{Conclusion}
In this paper, we propose new alignment methods that incorporate the concept of supervector in text-dependent speaker verification to systems based on neural networks.
The proposed alignment methods successfully encode the phrase and the identity in a supervector and are differentiable.
Furthermore, we present a novel optimization procedure to optimize the aAUC as an alternative to the triplet loss cost.
Both proposals have been evaluated in the text-dependent speaker verification database RSR2015 part I. 
%
As we show, training the system end-to-end to maximize
the AUC performance measure provides a better performance in
the detection task.


%



\section*{Acknowledgment}
This work has been supported by the Spanish Ministry of Economy and Competitiveness and the European Social Fund through the project TIN2017-85854-C4-1-R, by the Government of Aragon (Reference Group T36\_17R) and co-financed with Feder 2014-2020 "Building Europe from Aragon", and by Nuance Communications, Inc. We gratefully acknowledge the support of NVIDIA Corporation with the donation of the Titan Xp GPU used for this research.





\bibliographystyle{model1-num-names}
\bibliography{sample.bib}







\end{document}